# New Basis for the Standard Model for Electroweak Interactions

Mohammad Saleem
Institute for Basic Research
Palm Harbor, FL 34682, USA.

Muhammad Ali and Shaukat Ali
Theory Group, Department of Space Science
University of the Punjab, New Campus, Lahore. PAKISTAN.
e-mail: dms@lhr.paknet.com.pk

**Abstract**

The standard model for electroweak interactions uses the concepts of weak hypercharge and local gauge invariance of the Lagrangian density under the gauge group SU(2) x U(1). **Taylor** has remarked that U(1), being a multiply-connected group, in general, should have different coupling strengths for singlets and doublets. But in the development of the standard model, even for U(1), the coupling strength has been taken to be the same for all the multiplets without *a priori* justification. **Wilczek** has pointed out that the assignment of hypercharge values to left- and right-handed leptons and quarks is very peculiar. However, no solution has been provided. **Ellis** has also reservations for the rather bizarre set up of group representations and hypercharges. It has also been noticed that, in addition to the above weak points of the model, the generator of the group U(1) is changed at different stages during a single analysis. In this paper, a new basis is provided that avoids these peculiar and odd concepts and hypercharge assignments, and a satisfactory solution is given for the first time. This is done by generalising the expression for the transformation law for 4-vector potential in electromagnetism. It gives exactly the same results as obtained in the standard model for electroweak interactions.

The standard model for electroweak interactions [1] has been in excellent agreement with the enormous experimental data. However, several authors [2-6] have expressed their reservations about some basic assumptions in the model. It has been noticed that the assignment of weak hypercharge is peculiar and odd. The group U(1) is generated by weak hypercharges. By assigning peculiar values of these hypercharges to the left and right-handed electrons, etc. and by assuming the validity of the Gellman-Nishijima formula even for weak interactions, the results consistent with the experiment have been obtained [4]. The values $Y = -1$ and $Y = -2$ are assigned to the left-handed and the right-handed electrons. This is an assignment for which many physicists have reservations. Wilczek considers it as a very peculiar choice [2] and Ellis takes it as a bizarre assignment, and the only argument in its favour is that it gives the desired result. The same is true for quarks. The quantum numbers associated with the intrinsic properties do not change with the state of the particle but the weak hypercharge in the standard model has been allowed to do so. Moreover, in the standard model, U(1) is generated by hypercharge. It has been pointed out [6] that since the hypercharge varies from particle to particle, by stating that the group U(1) is generated by the hypercharge, actually the parametrisation of the group U(1) is changed at different stages. Instead of having the same single generator of group U(1) with a particular parametrisation throughout the analysis, in order to achieve the desired result, different multiples of the generator of U(1) have been used at different stages. It is therefore highly desirable that the standard model for electroweak interactions be built within a framework having a strong basis.

In the standard model for electroweak interactions, the symmetry group is $SU_L(2) \times U(1)_Y$. It has been pointed out by Taylor [5] that the group U(1) differs significantly from SU(2) in one very important aspect that relates to the topology of a symmetry group. The group SU(2) is simply-connected. Its manifold is compact. On the other hand, the group U(1) on which electromagnetic gauge invariance is based is different. Its manifold, the unit circle, is compact but it is *not* simply-connected. It is multiply-connected, because a closed path going a certain number of times around the circle cannot be deformed into one going a different number of times around. Therefore, U(1) behaves differently from the group SU(2). Hence, for U(1) that is mutiply-connected, there is no *a priori* reason why a different coupling constant should not be associated with each representation and also dependent upon the nature of the mutiplet. This characteristic of the group U(1) has constantly been ignored. We will take this into consideration and see that the masses of the exchange bosons and other characteristics of electroweak interactions can be derived without considering the weak hypercharge values and without the variation of the generator of U(1) at different stages during the analysis. Piecewise solutions based on various assumptions have also been suggested [6]. We will give here a comprehensive solution of the problem.

Let us see how can we achieve this. The left-handed electron and left-handed electron-neutrino form a doublet while the right-handed electron forms a singlet. SU(2) is coupled only to the left-handed doublet, the coupling constant being denoted by g. U(1) is coupled to the doublet as well as the singlet. Since U(1) is multiply-connected, there is *a priori* no justification for having the same coupling constant for every type of representation. The important point to notice is that in electromagnetism, the transformation law for the 4-vector $A_\mu$ is

$$A_\mu \rightarrow A_\mu' = A_\mu - \partial_\mu f(x).$$

This we agree to write as

$$A_\mu \rightarrow A_\mu' = A_\mu - 1/q\, \partial_\mu \theta(x),$$

where q is a constant and is the charge on the electron or is the coupling constant $g_s$ for U(1) singlet in electromagnetism. This is alright as long as we are considering electron as a singlet. However, this leads to difficulties when we go to doublets in the lepton sector and to singlets and doublets in the quark sector. It is this difficulty that led to the introduction of peculiar values to weak hypercharges. We therefore generalise it and write the transformation law for $A_\mu$ as

$$A_\mu \rightarrow A_\mu' = A_\mu - w/(ng_c)(-1)^a\, \partial_\mu \theta(x),$$

where n = 1 for singlet and 2 for doublet, w is the charge on the multiplet in terms of the electron charge and $g_c$ is the coupling constant of U(1) with respect to the multiplet. The value of a is 1 for a quark multiplet and 0 otherwise. Thus for the electromagnetic field for which we have n = 1, w = 1 and a = 0, the above equation reduces to the familiar form:

$$A_\mu' = A_\mu - 1/q\, \partial_\mu \theta(x)$$

It must be emphasized that in general the coupling constant for U(1) with respect to singlet or doublet changes with the multiplicity of the representation and the nature of the multiplicity but the transformation law for $A_\mu$ remains unchanged. The expression $(-1)^a\, w/ng_c$ therefore always remains equal to $1/g_s \equiv 1/q$.

Let us now make explicit calculations both for the lepton and quark sectors. We will confine ourselves to singlets and doublets. The left-handed parts of the e- and $\nu_e$ fields can be considered as forming a doublet of SU(2):

$$L \equiv \begin{pmatrix} \nu_{eL} \\ e_L \end{pmatrix}.$$

As the right-handed neutrinos do not exist in nature, the right-handed part of the neutrino field $\nu_{eR}$ would be zero: $\nu_{eR} = 0$ and the right-handed part of the electron-neutrino system would constitute a singlet $e_R$ only: $R \equiv e_R$. The other leptons can be treated in exactly the same way.

Let us now write down the leptonic part $\mathcal{L}_{lepton}$ of the Lagrangian density which is invariant under global $SU_L(2) \times U(1)$ transformations and for which spontaneous breaking of gauge symmetry can occur. It is

$$\mathcal{L}_{lepton} = i[\overline{L} \gamma^\mu \partial_\mu L + \overline{R} \gamma^\mu \partial_\mu R].$$

In order that the theory may be invariant under local $SU(2)_L \times U(1)$ transformations, the lepton part of the Lagrangian density should be modified to

$$\mathcal{L}_{lepton} = i[\overline{L} \gamma^\mu D_\mu^{(L)} L + \overline{R} \gamma^\mu D_\mu^{(R)} R],$$

where $D_\mu^{(L)}$ and $D_\mu^{(R)}$ are given by

$$D_\mu^{(L)} = \partial_\mu - ig'/2 \, G A_\mu - ig \, \mathbf{T}.\mathbf{W}_\mu$$
$$D_\mu^{(R)} = \partial_\mu - ig''/2 \, G A_\mu,$$

where $g'/2$ and $g''/2$ are the coupling constants of U(1) for the doublet and the singlet, respectively, and where G is the generator of the group and will be taken as $-1$ throughout the analysis. The coupling constant for the gauge group $SU_L(2)$ has been denoted by g. The important point is that although the coupling constant changes with the multiplicity of the representation in the case of U(1) as well as with its nature, the transformation law for $A_\mu$ must remain the same. The transformation law for $A_\mu$ in terms of the coupling constant $g'/2$ for the doublet is

$$A_\mu' = A_\mu - w/(ng_c)(-1)^a \, \partial_\mu \theta(x) = A_\mu - 1/g_s \, \partial_\mu \theta(x).$$

As $w = 1$, $n = 2$ and $a = 0$, this yields $g' = g_s$. Similarly, the transformation law for $A_\mu$ in terms of the coupling constant $g''/2$ for the singlet yields $g''/2 = g_s = g'$. Since $G = -1$, this leads to

$$D_\mu^{(L)} = \partial_\mu + ig'/2 \, A_\mu - ig \, \mathbf{T}.\mathbf{W}_\mu$$
$$D_\mu^{(R)} = \partial_\mu + ig' \, A_\mu.$$

The lepton mass term, $m\bar{e}e$, cannot be included in $\mathcal{L}_{\text{lepton}}$ because it is not invariant. Thus $\mathcal{L}_{\text{lepton}}$ may be expressed as

$$\mathcal{L}_{\text{lepton}} = i[\overline{L} \gamma^\mu (\partial_\mu - ig'/2\, GA_\mu - ig\, \mathbf{T}.\mathbf{W}_\mu)L + \overline{R}\gamma^\mu(\partial_\mu - ig'\, A_\mu)R].$$

The gauge part of the Lagrangian density is given by

$$\mathcal{L}_{\text{gauge}} = -1/4\, G_{\mu\nu}{}^j\, G^{\mu\nu j} - 1/4\, F_{\mu\nu} F^{\mu\nu},$$

where  $G_{\mu\nu}{}^j = (\partial_\mu W_\nu{}^j - \partial_\nu W_\mu{}^j) + g\, C_{\ell k}{}^j W_\mu{}^\ell W_\nu{}^k$

and  $F_{\mu\nu} = \partial_\mu A_\nu - \partial_\nu A_\mu$.

The masses for particles can be generated by introducing spontaneous symmetry breaking with scalar fields. For a spontaneous symmetry breaking which may make three of the four vector gauge bosons massive, and also give mass to the electron, we must use Higgs mechanism. That is, we should introduce a Higgs field, i.e., a scalar field with non-vanishing expectation value of the ground stat e which is not invariant under the gauge transformation. Since three massive and one massless exchange bosons are required, we need four independent scalar fields. The simplest possibility to accomplish this to introduce a complex weak isospin doublet of the Higgs scalars, one charged and one neutral:

$$\phi = \begin{pmatrix} \phi^+ \\ \phi^0 \end{pmatrix},$$

where $\phi^+$ and $\phi^0$ are scalar fields under Lorentz transformations. This complex scalar field transforms like a weak isospin doublet. The corresponding Lagrangian density which we denote by $\mathcal{L}_{\text{Higgs}}$ is

$$\mathcal{L}_{\text{Higgs}} = (D_\mu \phi)^\dagger (D^\mu \phi) - V(\phi^\dagger \phi),$$

where  $D_\mu \phi = (\partial_\mu - ig_H/2\, GA_\mu - ig\, \mathbf{T}.\mathbf{W}_\mu)\phi$,

$V(\phi^\dagger \phi) = \mu^2 \phi^\dagger \phi + \lambda (\phi^\dagger \phi)^2$,

and $g_H/2$ is the coupling constant for the scalar doublet. By virtue of the fact that the transformation law for $A_\mu$ has to remain the same and that the charge on the scalar doublet is 1, we have $g_H = -g'$. Since $G = -1$, we have

$$D_\mu = (\partial_\mu - ig'/2\, A_\mu - ig\, \mathbf{T}.\mathbf{W}_\mu).$$

We now consider the consequences of the spontaneous symmetry breaking. We have seen that the part of the Lagrangian density which involves Higgs scalars is given by

$$\mathcal{L}_{\text{Higgs}} = (D_\mu \phi)^\dagger (D^\mu \phi) - V(\phi^\dagger \phi).$$

The detailed analysis yields

$$\mathcal{L}_{\text{Higgs}} = 1/2[(\partial_\mu \eta)(\partial^\mu \eta)$$

$$+ 1/4 \, (v + \eta)^2 \{(-gW_\mu^3 + g' A_\mu)(-gW^{\mu 3} + g' A^\mu)$$
$$+ g^2 \, (W_\mu^1 W^{\mu 1} + W_\mu^2 W^{\mu 2})\}]$$
$$- \mu^2/2 (v^2 + \eta^2 + 2 v \eta)$$
$$+ \mu^2/4v^2 (v^4 + \eta^4 + 6 v^2 \eta^2 + 4 v^3 \eta + 4 v \eta^3).$$

The presence of the term $\partial_\mu \eta \partial^\mu \eta$ in the expression for $\mathcal{L}_{Higgs}$ shows that the field $\eta$ exists. The coefficient of $-1/2 \, \eta^2$ with no coupling with other fields gives the square of the mass of the quantum of this field:

$$m^2_\eta = -2\mu^2.$$

This particle is termed as the **Higgs boson**. Since the parameter $\mu^2$ is free, except that it should be negative, the mass of the Higgs boson is not determined.

Proceeding in the usual way, we get exactly the same value for the masses of the exchange particles as in the conventional standard model.

Let us next consider the quark sector. For simplicity, we will confine ourselves to u and d quarks. Then the left-handed components of u and d quarks form an isospin doublet of $SU_L(2)$:

$$L_{quark} \equiv \begin{pmatrix} u_L \\ d_L \end{pmatrix}.$$

The right-handed components on the other hand are singlets. We then have

$$\mathcal{L}_{quark} = i[\, \overline{L}_{quark} \gamma^\mu (\partial_\mu - ig'_h/2 \, GA_\mu - ig \, \mathbf{T.W}_\mu) L_{quark}$$
$$+ \bar{u}_R \gamma^\mu (\partial_\mu - ig''_h/2 \, GA_\mu) u_R$$
$$+ \bar{d}_R \gamma^\mu (\partial_\mu - ig'''_h/2 \, GA_\mu) d_R] + \text{Yukawa terms.}$$

Proceeding as before, we have

$$-(1/3)/(2g'_h/2) = 1/g'$$

or $\quad g'_h = -g'/3.$

Similarly, we obtain

$\quad g''_h = -4g'/3$

and $\quad g'''_h = 2g'/3.$

Substituting these expressions in the equation for the Lagrangian density for the quark sector and noting that $G = -1$, we have

$$\mathcal{L}_{quark} = i[\, \overline{L}_{quark} \gamma^\mu (\partial_\mu - ig'/2 \times 1/3 \, A_\mu - ig \, \mathbf{T.W}_\mu) L_{quark}$$
$$+ \bar{u}_R \gamma^\mu (\partial_\mu - ig'/2 \times 4/3 A_\mu) u_R$$
$$+ \bar{d}_R \gamma^\mu (\partial_\mu + ig'/2 \times 2/3 \, A_\mu) d_R] + \text{Yukawa terms.}$$

This is exactly the same as for the conventional standard model. Hence, this will yield the same results. We conclude that the concept of weak hypercharge with peculiar values for the left- and right-handed electrons and quarks obtained by assuming the Gellman-Nishijima formula for weak interactions can be replaced by using a logical and consistent conceptual framework based on the fact that U(1) is a multiply-connected group and by generalising the expression for the transformation law for 4-potential in electromagnetism. This also avoids the variation of the generator of U(1) at various stages during the analysis.